\documentclass{article}

\usepackage[preprint, nonatbib]{nips_2018}

\usepackage[utf8]{inputenc} 
\usepackage[T1]{fontenc}    
\usepackage{hyperref}       
\usepackage{url}            
\usepackage{booktabs}       
\usepackage{amsfonts}       
\usepackage{nicefrac}       
\usepackage{microtype}      
\usepackage{amsthm}
\usepackage{thmtools}
\usepackage{thm-restate}
\usepackage{amssymb}
\usepackage{wrapfig}
\usepackage{csquotes}
\usepackage[all,cmtip]{xy}
\usepackage{graphicx}
\usepackage{caption}
\usepackage{subcaption}
\usepackage{stmaryrd}
\usepackage{multicol}
\usepackage{mathtools}
\usepackage[english]{babel}
\usepackage{dsfont}
\usepackage{floatrow}
\newfloatcommand{capbtabbox}{table}[][\FBwidth]
\graphicspath{{images/}}

\theoremstyle{definition}

\theoremstyle{remark}

\makeatletter
\numberwithin{equation}{section}
\numberwithin{figure}{section}

\makeatother

\newcommand{\R}{\mathbb{R}}

\newcommand{\abs}[1]{\left|#1\right|}

\renewcommand{\c}[1]{\mathcal{#1}}

\newcommand{\norm}[1]{\left\|#1\right\|}

\DeclarePairedDelimiterX{\infdivx}[2]{(}{)}{%
	#1\;\delimsize\|\;#2%
}

\usepackage{babel}
\usepackage{tikz}

\definecolor{mydarkblue}{rgb}{0,0.08,0.45}
\definecolor{myfavblue}{rgb}{0.1176, 0.392, 1.0}
\hypersetup{
    colorlinks=true,
    linkcolor=mydarkblue,
    citecolor=mydarkblue,
    filecolor=mydarkblue,
    urlcolor=mydarkblue}

\usetikzlibrary{arrows,calc}

\usepackage{algorithm}
\usepackage[noend]{algpseudocode}

\algrenewcommand\algorithmicindent{1.0em}%
\algrenewcommand\textproc{}

\newcommand{\prototype}{prototype}

\newcommand{\latents}{vectors}

\newcommand{\uvec}{user\_vector}
\newcommand{\ivec}{item\_vector}
\newcommand{\cuvec}{{\color{blue}{\uvec}}}
\newcommand{\civec}{{\color{red}{\ivec}}}

\newcommand{\depth}{{\texttt{depth}}}

\newcommand{\el}{{\mathbf{EL}}}
\newcommand{\md}{{\mathbf{MD}}}
\newcommand{\pu}{{\mathbf{P_u}}}
\newcommand{\pv}{{\mathbf{P_v}}}

\title{Scalable Recommender Systems\\ through Recursive Evidence
Chains}

\author{
  Elias Tragas \thanks{Part of this work done while author was a student at University of Toronto.} \\
  Snapchat, Inc.\\
  \texttt{etragas@gmail.com} \\
    \And
    Calvin Luo \\
    University of Toronto \\
    \texttt{exetercluo@gmail.com} \\
  \And
  Maxime Gazeau \\
  Borealis AI \\
  \texttt{maxime.gazeau@borealisai.com} \\
  \And
  Kevin Luk \\
  Borealis AI \\
  \texttt{kevin.luk@borealisai.com} \\
  \And
  David Duvenaud \\
  University of Toronto \\
  Vector Institute \\
  \texttt{duvenaud@cs.toronto.edu} \\
}

\begin{document}

\maketitle

\begin{abstract}
Recommender systems can be formulated as a matrix completion problem, predicting ratings from user and item parameter vectors.
Optimizing these parameters by subsampling data becomes difficult as the number of users and items grows.
We develop a novel approach to generate all latent variables on demand from the ratings matrix itself and a fixed pool of parameters.
We estimate missing ratings using chains of evidence that link them to a small set of prototypical users and items.
Our model automatically addresses the cold-start and online learning problems by combining information across both users and items.
We investigate the scaling behavior of this model, and demonstrate competitive results with respect to current matrix factorization techniques in terms of accuracy and convergence speed. 
\end{abstract}

\section{Introduction} \label{sec:intro}

The central aim of model-based collaborative-filtering methods is to predict a user's rating of an item from a small number of recorded preferences in the system. An effective approach towards this problem is to formulate it as a matrix factorization problem. One can approximate a ratings matrix $\c{R}\in\R^{M\times N}$ as a low-rank factorization $\c{R}\approx\c{U}\c{V}^{\top}$, where $\c{U}\in\R^{M\times K}$, $\c{V}\in\R^{N\times K}$ and $K\ll\min(M,N)$~\cite{srebro2003weighted,rennie2005fast}.
The $K$-dimensional rows $\c{U}_i$ and $\c{V}_j$ of $\c{U}$ and $\c{V}$ are commonly referred to as the latent user and item vectors. 
Under this framework, each of the ratings entries $\c{R}_{ij}$ is approximated by the inner product $\c{U}_i\c{V}_j^{\top}$.

\begin{figure}
    \centering%
    \begin{subfigure}[t]{.33\textwidth}%
    \centering%
    \begin{tikzpicture} {
    \begin{scope}[every node/.style={draw,thin,align=center}]
    \node (V) at (0,1.1) [anchor=south west,draw,thick,minimum width=2cm,minimum height=0cm] {$\c{V}$};
    \node (U) at (-.1,1)   [anchor=north east,draw,thick,minimum width=0cm,minimum height=2cm] {$\c{U}$};
    \node (R) at (1,0) [draw,thick,minimum width=2cm,minimum height=2cm] {$\c{R}$};
    \draw[line width=0cm,color=white] (-1,-.99) -- (2.1,-.99) ;    
    \end{scope}
}\end{tikzpicture}
\caption{Latent factor models}
    \end{subfigure}%
\begin{subfigure}[t]{.33\textwidth}
    \centering%
    \begin{tikzpicture} {
    \begin{scope}[every node/.style={thin,align=center}]
    \node (V) at (0,1.1) [anchor=south west,draw,thick,minimum width=2cm,minimum height=0cm] {$\c{V}$};
    \node (U) at (-.1,1)   [anchor=north east,draw,dash pattern=on 2pt off 2pt,minimum width=0cm,minimum height=2cm] {$\hat{\c{U}}$};
    \node (R) at (1,0) [draw,thick,minimum width=2cm,minimum height=2cm] {$\c{R}$};
    \node (elipsesV) at (-0.4,-1.2) [rotate=90,line width=0] {$...$};
    \node (elipsesR) at (1,-1.2) [rotate=90,line width=0] {$...$};
    \draw[line width=.1cm,color=white] (-1.2,-.99) -- (2.1,-.99) ;
    \end{scope}
}\end{tikzpicture}    
    \caption{Rowless factor model}
    \label{fig:rowless}
    \end{subfigure}%
\begin{subfigure}[t]{.33\textwidth}
    \centering%
    \begin{tikzpicture} {
    \begin{scope}[every node/.style={thin,align=center}]
    \node (Vhat) at (0,1.1)   [anchor=south west,draw,dash pattern=on 2pt off 2pt,minimum width=2cm,minimum height=.55cm] {$\hat{\c{V}}$};
    \node (V)    at (0,1.1)  [anchor=south west,draw,thick,minimum width=.75cm,minimum height=.55cm] {$\c{V}$};
    \node (Uhat) at (-.1,1)   [anchor=north east,draw,dash pattern=on 2pt off 2pt,minimum width=0.55cm,minimum height=2cm] {$\hat{\c{U}}$};
    \node (U)    at (-.1,1)   [anchor=north east,draw,thick,minimum width=0.55cm,minimum height=.75cm] {$\c{U}$};
    \node (R) at (1,0) [draw,thick,minimum width=2cm,minimum height=2cm] {$\c{R}$};
    \node (elipsesV) at (-0.4,-1.1) [rotate=90,line width=0] {$...$};
    \node (elipsesRD) at (1,-1.2) [rotate=90,line width=0] {$...$};
    \node (elipsesRDiag) at (2.1,-1.2) [rotate=135,line width=0] {$...$};
    \node (elipsesRR) at (2.2,0) [rotate=0,line width=0] {$...$};
    \node (elipsesV) at (2.2,1.35) [rotate=0,line width=0] {$...$};
    \draw[line width=.1cm,color=white] (-1.2,-.99) -- (2.1,-.99) ;
    \draw[line width=.1cm,color=white] (2,-.99) -- (2,1.57) ;
    \end{scope}
}\end{tikzpicture}    
    \caption{Proposed method}
    \label{fig:us}
    \end{subfigure}%
\caption{Dotted lines denote generated embeddings, as opposed to stored in memory. Ellipses show the direction in which the data can scale without having to add new parameters.}
\label{fig:three approaches}
\end{figure}
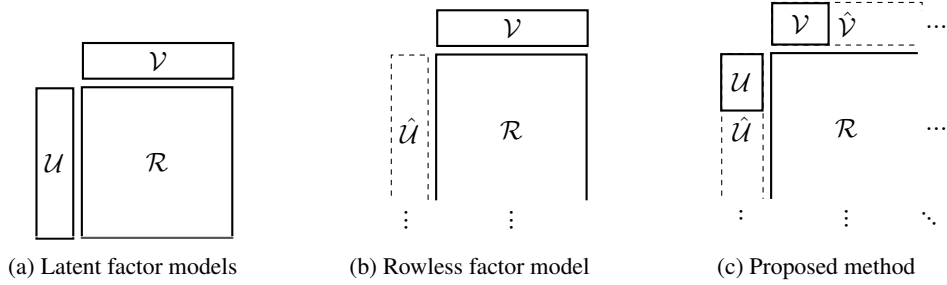

A major drawback to these methods is that the number of parameters to be optimized grows with the number of users and items.
Because the training loss is a coupled function of all of these parameters, stochastic optimization through data mini-batches resembles a form of blockwise coordinate optimization.
We conjecture that there is room for improvement in the stochastic optimization of large models of this form, by directly enforcing a coupling between parameters of similar rows and columns. 


In this paper, we propose our Recursive Evidence Chains (REC) algorithm. The central idea to our algorithm is that we do not store the entire latent matrices $\c{U}$ and $\c{V}$ but rather store only a very small subset of it (the prototypes) and then learn a function using neural networks to recursively generate the latent representations for non-prototypical users and items on-demand.

Another challenge for collaborative-filtering methods is the classic \enquote{cold-start} problem, where new users have few ratings and the system has limited information on the preferences of such users.
A side-benefit of applying recursive chains to generate latent representations on-demand is that the coupling of parameters between users provides a natural form of information-sharing, or regularization.

\section{Background}
\label{sec:background}


In this section we introduce the notion of rowless and columnless matrix factorization techniques, as well as explore how they relate to coupled parameter optimization and online learning. In the next section, we will show how our proposed algorithm REC combines and generalizes these approaches.

\paragraph{Rowless collaborative filtering}\label{sec:rowless}

Instead of explicitly storing an embedding for every row and column in a matrix, rowless methods generate row embeddings on-demand. They estimate each embedding as a function of the specific row’s rated item embeddings in conjunction with the corresponding ratings themselves.
One such function could be a neural net which maps any given (item embedding, rating) pair to a user embedding. Let $f_\phi$ be such a net, then we can generate some user embedding $\widehat{\c{U}}_i$ by taking the average of $f_\phi$ over all of user $i$'s rated items, which we denote by $\Omega_i$. In precise mathematical terms, we have
\begin{equation}\label{row_less}
        \hat{\c{U}}_i =  \frac{1}{\abs{\Omega_i}} \sum_{j \in \Omega_i}
        f_{\varphi}\left(\c{R}_{ij},  \c{V}_{j} \right)
\end{equation} where $\c{V}_{j}$ denotes the embedding for item $j$ and $\c{R}_{ij}$ denotes the user $i$'s rating for item $j$. After generating $\hat{\c{U}}_i$, we can make a prediction by setting  $\hat{\c{R}}_{ij} = \hat{\c{U}}_i\c{V}_j^{\top}$.

\paragraph{Properties of rowless methods}
Rowless methods perform well in online settings~\cite{das2016chains,verga2016generalizing}, since they can handle an infinite amount of novel rows without requiring retraining. In addition, they have $O(NK)$ parameter complexity, since row embeddings are automatically generated from item embeddings. We will show in our experiments in Section~\ref{sec:experiments} that using an aggregation function to couple multiple embeddings can improve the rate of convergence for mini-batch gradient based optimization methods. 

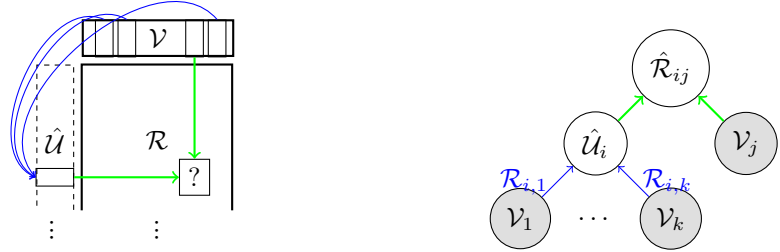
\begin{figure}[h]
\label{row-less}
\centering
\begin{subfigure}[t]{.5\textwidth}
    \centering
    \begin{tikzpicture} {
    \begin{scope}[every node/.style={thin,align=center}]
    \node (V) at (0,1.1) [anchor=south west,draw,thick,minimum width=2cm,minimum height=0cm] {$\c{V}$};
    \node (U) at (-.1,1)   [anchor=north east,draw,dash pattern=on 2pt off 2pt,minimum width=0cm,minimum height=2cm] {$\hat{\c{U}}$};
    \node (R) at (1,0) [draw,thick,minimum width=2cm,minimum height=2cm] {$\c{R}$};
    \node (elipsesV) at (-0.4,-1.2) [rotate=90,line width=0] {$...$};
    \node (elipsesR) at (1,-1.2) [rotate=90,line width=0] {$...$};
    \node (que) at (1.5,-.5) [draw] {$?$};
    \node (mj) at(1.5,1.095)  [anchor = south,rectangle,minimum height=.49cm,draw] {};
    \node (uj) at(-.1,-.5)  [anchor = east,rectangle,minimum width=.5cm,draw] {};
    \node (m1) at(.3,1.095)  [anchor = south,rectangle,minimum height=.49cm,draw] {};
    \node (m2) at(.6,1.095)  [anchor = south,rectangle,minimum height=.49cm,draw] {};
    \node (m3) at(1.8,1.095)  [anchor = south,rectangle,minimum height=.49cm,draw] {};
    
    \draw[color=green,thick] (mj) edge[->] (que)
                       (uj) edge[->] (que);
   \draw[color=blue,thin]
                       (m1.north) edge[->,bend right=90] (uj.west)
                       (m2.north) edge[->,bend right=90] (uj.west)
                       (m3.north) edge[->,bend right=90] (uj.west);
    \draw[line width=.1cm,color=white] (-1.2,-.99) -- (2.1,-.99) ;
    \end{scope}
}\end{tikzpicture}    
    \caption{Information flow for a rowless factor model}
    \label{fig:rowless flow}
    \end{subfigure}%
    \begin{subfigure}[t]{.5 \textwidth}
        \centering
\begin{tikzpicture} {
\begin{scope}[every node/.style={thin,align=center}]

  \node (rating) at (2,1) [circle,draw] {$\hat{\c{R}}_{ij}$};
    \node (vi) at (1,0) [circle,draw] {$\hat{\c{U}}_i$};
    \node (ui) at (3,0) [circle,draw,fill=gray!25] {$\c{V}_j$};
    \node (u1) at (0, -1) [circle,draw,fill=gray!25] { $\c{V}_1$};
    \node (dots) at (1, -1) [circle] {$\dots$};
    \node (uk) at (2, -1) [circle,draw,fill=gray!25] { $\c{V}_k$};
    \draw   
            (u1) edge[->,color=blue,thin] node[left]{$\c{R}_{i,1}$} (vi)
            (uk) edge[->,color=blue,thin] node[right]{$\c{R}_{i,k}$} (vi)
            (ui) edge[->,color=green,thick] (rating)
            (vi) edge[->,color=green,thick] (rating);
\end{scope}
}\end{tikzpicture}    
    \caption{Graphical model of making predictions in a rowless factor model.}
    \label{fig:rowless predict}
\end{subfigure}%

\caption{Two alternative representations of information flow in a rowless model. We see that multiple item embeddings and their ratings are used to define $\hat{\c{U}}_i$. In turn, $\hat{\c{U}}_i$ and ${\c{V}}_j$ are used to define $\hat{\c{R}}_{ij}$.}
\end{figure}




\paragraph{Columnless collaborative filtering} 
By making the necessary modifications, we can imagine a  columnless model:
\begin{equation}\label{column_less}
        \hat{\c{V}}_j =  \frac{1}{\abs{\overline{\Omega}_j}} \sum_{i \in \overline{\Omega}_j}
        f_{\psi}\left(\c{R}_{ij}, \hat{\c{U}}_i \right)
\end{equation}
where $\overline{\Omega}_j$ denotes the rows in which a rating for item $j$ exists, and $f_\psi$ represents a neural network that maps a given (user embedding, rating) pair to an item embedding.  Analogous to the rowless case, columnless methods have $O(MK)$ space complexity.

\paragraph{Combining rowless and columnless methods}
Rowless and columnless methods rely on $O(NK)$ and $O(MK)$ parameters respectively, where $K$ is the latent factor size. In contrast, Singular Value Decomposition (SVD) has parameter complexity $O(NK+MK)$. Our proposed algorithm REC, which we present the next section, leverages recursion to achieve $O(K)$ parameter scaling with respect to the dataset size.

\section{Recursive Evidence Chains} \label{sec:REC}


In this section we introduce a general framework for combining both rowless and columnless matrix factorization.
Instead of allocating full embeddings for only columns or only rows, we pick a constant number of prototype users $\pu$ and prototype items $\pv$ such that $\abs{\pu} \ll M$ and $\abs{\pv} \ll N$. We select our prototypes to be the users and items with the most ratings. To aid with visualization, we sort our matrices such that the $\pu$ users are the first $\abs{\pu}$ rows of the matrix, and similarly so for items.

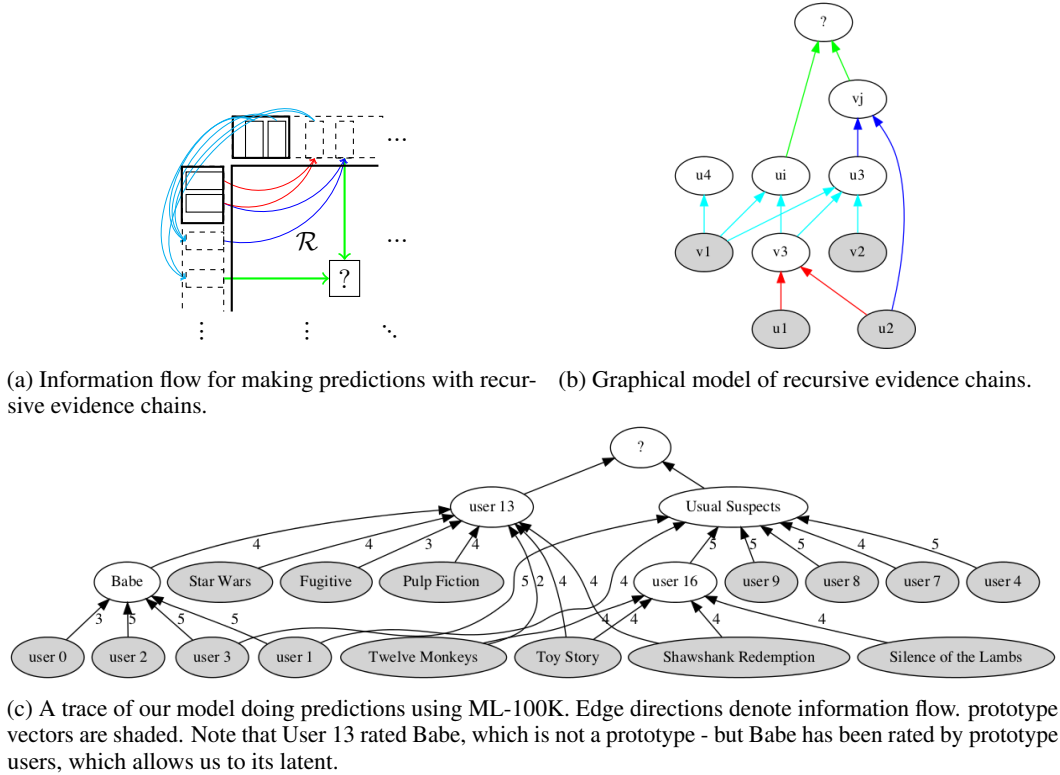
\begin{figure}[h]
    \begin{subfigure}[t]{.5\textwidth}
    \centering
    \begin{tikzpicture} {
    \begin{scope}[every node/.style={thin,align=center}]
    \node (U) at (0,1.1) [anchor=south west,draw,dash pattern=on 2pt off 2pt,minimum width=2cm,minimum height=.55cm] {};
    \node (U) at (0,1.09) [anchor=south west,draw,thick,minimum width=.75cm,minimum height=.55cm] {};
    \node (que) at (1.5,-.5) [draw] {$?$};
    \node (mj) at(1.5,1.095)  [anchor = south,rectangle,dash pattern=on 2pt off 2pt,minimum height=.49cm,draw] {};
    \node (uj) at(-.1,-.5)  [anchor = east,rectangle,dash pattern=on 2pt off 2pt,minimum width=.5cm,draw] {};
    \node (m1) at(.3,1.095) [anchor = south,rectangle,minimum height=.49cm,draw] {};
    \node (m2) at(.6,1.095) [anchor = south,rectangle,minimum height=.49cm,draw] {};
    \node (m3) at(1.1,1.095) [anchor = south,rectangle,minimum height=.49cm,dash pattern=on 2pt off 2pt,draw] {};
    \node (u1) at(-.1,.5)  [anchor = east,rectangle,minimum width=.5cm,draw] {};
    \node (u2) at(-.1,.8)  [anchor = east,rectangle,minimum width=.5cm,draw] {};
    \node (u3) at(-.1,0)  [anchor = east,rectangle,minimum width=.5cm,dash pattern=on 2pt off 2pt,draw] {};
    \draw[color=green,thick] (mj) edge[->] (que)
                       (uj) edge[->] (que);
    \draw[color=blue,very thin]
                       (u1.east) edge[->,bend right=45] (mj.south)
                       (u3.east) edge[->,bend right=45] (mj.south);
    
    \draw[color=red,very thin] 
                       (u1.east) edge[->,bend right=45] (m3.south)
                       (u2.east) edge[->,bend right=45] (m3.south);
    \draw[color=cyan,very thin]
                       (m1.north) edge[->,bend right=90] (u3.west)
                       (m2.north) edge[->,bend right=90] (u3.west)
                       (m3.north) edge[->,bend right=90] (u3.west)
                       (m1.north) edge[->,bend right=90] (uj.west)
                       (m3.north) edge[->,bend right=90] (uj.west);
    \node (V) at (-.1,1)   [anchor=north east,draw,dash pattern=on 2pt off 2pt,minimum width=0.55cm,minimum height=2cm] {};
    \node (V) at (-.1,1)   [anchor=north east,draw,thick,minimum width=0.55cm,minimum height=.75cm] {};
    \node (R) at (1,0) [draw,thick,minimum width=2cm,minimum height=2cm] {$\c{R}$};
    \node (elipsesV) at (-0.4,-1.2) [rotate=90,line width=0] {$...$};
    \node (elipsesRD) at (1,-1.2) [rotate=90,line width=0] {$...$};
    \node (elipsesRDiag) at (2.1,-1.2) [rotate=135,line width=0] {$...$};
    \node (elipsesRR) at (2.2,0) [rotate=0,line width=0] {$...$};
    \node (elipsesV) at (2.2,1.35) [rotate=0,line width=0] {$...$};
    \draw[line width=.1cm,color=white] (-1.2,-.99) -- (2.1,-.99) ;
    \draw[line width=.1cm,color=white] (2,-.99) -- (2,1.57) ;
    \end{scope}
}\end{tikzpicture}    
   \captionsetup[figure]{width=.8\linewidth,justification=centering}

    \captionof{figure}{Information flow for making predictions with recursive evidence chains.}
    \label{fig:rec flow}
    \end{subfigure}%
\begin{subfigure}[t]{.5\textwidth}
    \centering
\includegraphics[scale=.3]{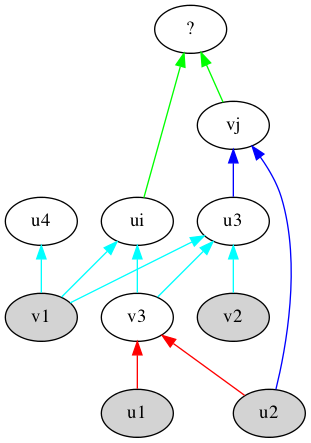}    
   \captionsetup[figure]{width=.8\linewidth,justification=centering}

\captionof{figure}{Graphical model of recursive evidence chains.}
\label{fig:recgraph}
\end{subfigure}%

\begin{subfigure}{\textwidth}
    \includegraphics[width=\columnwidth]{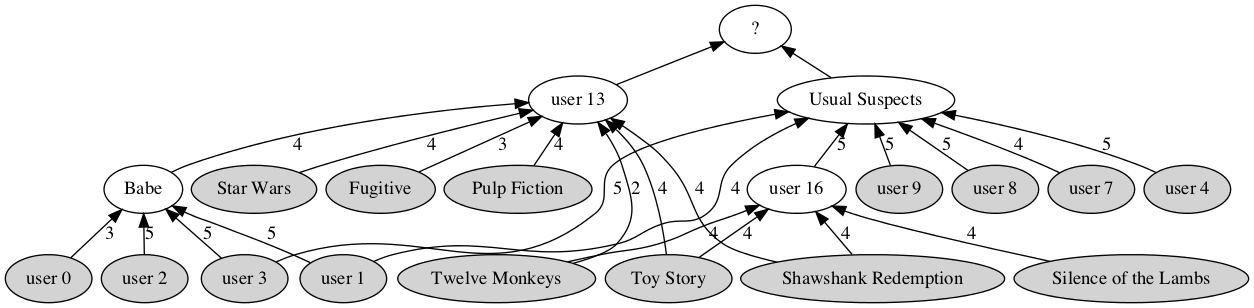}
    
    \captionof{figure}{A trace of our model doing predictions using ML-100K. Edge directions denote information flow.
    \prototype{} \latents{} are shaded. Note that User 13 rated Babe, which is not a \prototype{} - but Babe has been rated by \prototype{} users, which allows us to  its latent.}
\end{subfigure}%
\caption{Recursive prediction in REC}
\end{figure}

Each prototype user and item receives an embedding, while all non-prototypical users $u_i$ and non-prototypical items $v_j$ are predicted on-demand.  Intuitively, because our method is rowless, we are able to predict each missing $u_i$ embedding as a function of the ratings of that user and the embeddings of each item the user rated. Since our method is also columnless, we predict each missing $v_j$ as a function of the ratings of that item and the embeddings of each user who rated that item.
This introduces a recursion until we reach the prototypical users and items.
An example of this type of recursive structure is shown in Figures~\ref{fig:rec flow} and~\ref{fig:recgraph}. 


\subsection{Predicting latent factors in REC}
We denote by $f_{\varphi}$ and $f_{\psi}$ two feed-forward networks
parametrized respectively by $\varphi \in \R^d$ and $\psi \in \R^p$.
The former maps an item latent factor and rating pair to a user latent factor,
while the latter maps a user latent factor and rating pair to an item latent
factor. 


When collecting evidence to generate an embedding, we decompose the problem into whether or not our evidence stems from a prototypical user or item.
For users, we define the latent factors as
\begin{equation}\label{eq:u}
    {\color{blue}  \c{U}_i} = \begin{cases}
u_i & \mbox{if } i\in \pu \\
        \hat{u}_i =  \frac{1}{\abs{\Omega_i}} \sum_{\ell \in \Omega_i}
        f_{\varphi}\left(\c{R}_{i\ell}, {\color{red} \c{V}_{\ell} } \right) & \mbox{otherwise}.
\end{cases}
\end{equation}
where we recall that $\Omega_i$ is the set of rated items by the $i$-th user.
Similarly, for items, we have the function $\c{V}_j$:
\begin{equation}\label{eq:v}
    {\color{red} \c{V}_j} = \begin{cases}
        v_j & \mbox{if } j\in\pv \\
        \hat{v}_j =  \frac{1}{\abs{\overline{\Omega}_j}} \sum_{\ell \in
        \overline{\Omega}_j}
        f_{\psi}\left(\c{R}_{\ell j}, {\color{blue} \c{U}_{\ell}}  \right)& \mbox{otherwise},
\end{cases}
\end{equation}
 where $\overline{\Omega}_j$ is the set of users that gave a rating for item $j$. 
 
With this simple formulation, it is almost impossible for an embedding generation step of REC to finish.
If at least one rating $\c{R}_{ij}$ is shared between a non-prototype row and non-prototype column, attempting to generate their embeddings will cause an infinite loop, since each predicted embedding $\hat{u}_i$, $\hat{v}_j$ will request the other's value. To address this, we introduce a Max Depth $(\md)$ constant. If a recursive call has depth greater than or equal to $\md$ in the stack, and the requested embedding is not a prototype, we return None and ignore that embedding's value in the summation. This guarantees that REC will always terminate.

\paragraph{Predictions with REC}
For a given rating $\c{R}_{ij}$, we generate our predicted ratings $\hat{\c{R}}_{ij}$ by generating both $\c{U}_i$ and $\c{V}_j$ and then setting $\hat{\c{R}}_{ij} =  \c{U}_i\c{V}_j^T$. It is possible for one or both of our generated embeddings to be undefined. This can occur if the shortest path to the prototypes is longer than our given $\md$. In this case, we return the mean of the dataset. 

\paragraph{Training REC}
Given the above, we are able to train REC end-to-end using SGD. We can jointly optimize REC's set of parameters: $\Theta =  \left\{u, v, \phi, \psi \right\}$ where $u$ are the user prototype embeddings, $v$ the item prototype embeddings and $\phi$, $\psi$ the parameters of our generator nets. While the definition of our latent feature vectors $(\c{U}, \c{V})$ rely on piece-wise functions, they are sub-differentiable and thus easy to optimize in a framework which supports automatic differentiation such as PyTorch or Tensorflow. We use the standard matrix factorization loss function, with regularization terms on the magnitude of our prototype embeddings and generator net parameters.
\begin{equation}\label{eq:REC}
    \min_{\Theta} L(\Theta) = \sum_{(i,j) \in \Omega}
     \left(\c{R}_{ij}- \c{U}_i\c{V}_j^T \right)^2 + \lambda \norm{u}^2_F
     + \lambda \norm{v}^2_F + \lambda \left( \norm{\varphi}^2 + \norm{\psi}^2 \right).
\end{equation}

\begin{figure}[h]\label{pseudo_code_rec}
\begin{algorithm}[H]
\caption{Prediction in REC with Max Depth}
\centering
\begin{algorithmic}[1]
\Procedure{\texttt{recursive\_predict\_rating}}{$i, j$}
    \State $\texttt{$\bar{u}_i$} \gets \texttt{\cuvec{}}(i, \depth = 0)$
    \State $\texttt{$\bar{v}_j$} \gets \texttt{\civec{}}(j, \depth = 0)$
    \If $\texttt{$\bar{u}_i$ == None} \;\; ||\;\; \texttt{$\bar{v}_j$ == None}$
    \State \Return $\texttt{dataset\_mean}$
    \EndIf
    \State \Return $\bar{u}_i\bar{v}_j^T$
\EndProcedure
\end{algorithmic}
\begin{multicols}{2}%
\begin{algorithmic}[1]%
\Procedure{$\texttt{\cuvec{}}$}{$i, \depth$}
\If {$\texttt{is\_prototype\_user(i)}$}: 
\State \Return $\texttt{embedding\_for\_user(i)}$
\EndIf
\If {$\depth \geq \texttt{max\_depth}:$}
\State \Return $\texttt{None}$
\EndIf
\State $\texttt{user\_embeddings} = []$
\For {$j \in \texttt{items\_rated\_by\_user(i)}:$}
\State $\bar{v}_j \gets \texttt{\civec}(j, \depth + 1)$
\If \texttt{$\bar{v}_j$ == None}:
\State $\texttt{continue}$
\EndIf
\State \texttt{user\_embeddings.append($f_{\varphi}(\bar{v}_j,\c{R}_{ij})$)}
\EndFor
\If {$\texttt{user\_embeddings} == []:$}
\State \Return $\texttt{None}$
\EndIf
\State \Return \texttt{mean(user\_embeddings)}
\EndProcedure
\end{algorithmic}
\columnbreak%
\begin{algorithmic}[1]%
\Procedure{$\texttt{\civec{}}$}{$j, \depth$}
\If {$\texttt{is\_prototype\_item(j)}$}: 
\State \Return $\texttt{embedding\_for\_item(j)}$
\EndIf
\If {$\depth \geq \texttt{max\_depth}:$}
\State \Return $\texttt{None}$
\EndIf
\State $\texttt{item\_embeddings} = []$
\For {$i \in \texttt{users\_rated\_by\_item(j)}:$}
\State $\bar{u}_i \gets \texttt{\cuvec}(i, \depth + 1)$
\If \texttt{$\bar{u}_i$ == None}:
\State $\texttt{continue}$
\EndIf
\State \texttt{item\_embeddings.append($f_{\psi}(\bar{u}_i,\c{R}_{ij})$)}
\EndFor
\If {$\texttt{item\_embeddings} == []:$}
\State \Return $\texttt{None}$
\EndIf

\State \Return \texttt{mean(item\_embeddings)}
\EndProcedure
\end{algorithmic}
\end{multicols}
\end{algorithm}
\caption{Pseudocode for REC with Max Depth. $f_\phi$ and $f_\psi$ here denotes the neural networks used to generate our user and item vectors respectively.}
\end{figure}

\section{Complexity controls} \label{sec:complexity}

\begin{figure}[h]
    \centering%
    \begin{subfigure}[t]{.5\textwidth}%
    \centering%
      \includegraphics[width=\linewidth]{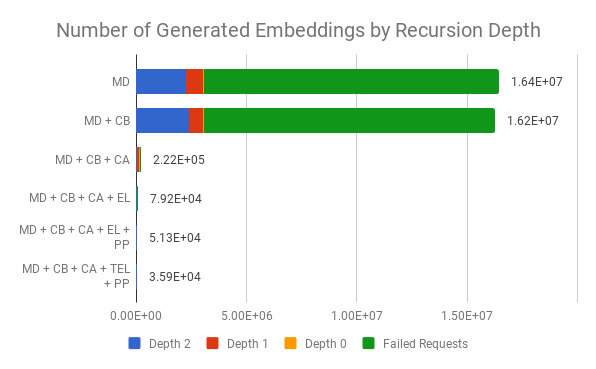}
    \label{fig:us}
    \end{subfigure}%
\begin{subfigure}[t]{.5\textwidth}
    \centering%
      \includegraphics[width=\linewidth]{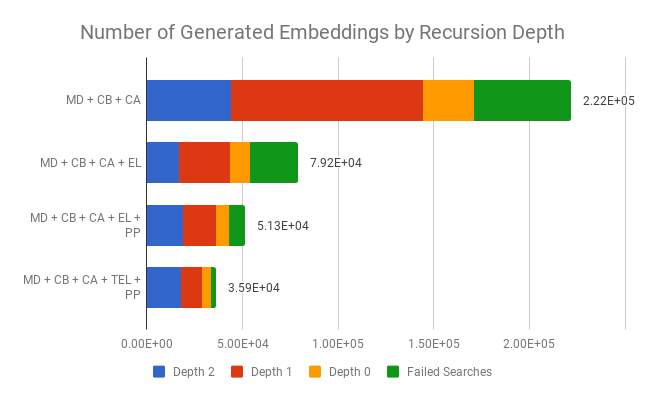}
    \label{fig:us}
    \end{subfigure}%
\caption{Embeddings generated for a mini-batch of size 10 on ML-100K. Max Depth is set to 2 for all experiments. On the left, we see the behaviour for 6 different implementations of REC using various complexity controls. On the right, we add the cache and note its ability to reduce the number of failed requests.}
\label{fig:complexityControl}
\end{figure}

While REC with an assigned Max Depth can converge, it suffers from wasted computation. In this section we introduce complexity controls to minimize the amount of computation done by REC, while still retaining enough information to accurately predict a given rating. When we combine these complexity controls, the amount of computation required is reduced by 3 orders of magnitude compared to an implementation using only Max Depth. The impact of our complexity controls can be seen in Figure~\ref{fig:complexityControl}.

\textbf{Cycle Blocking (CB)} We begin by eliminating cycles from the computation. Before using some embedding $\hat{v}_j$ as evidence to generate some embedding $\hat{u}_i$, we first check if $\hat{u}_i$ is already acting as evidence for $\hat{v}_j$ earlier in the call stack. If so, we ignore it.

\textbf{Caching (CA)}
When generating some embedding $\hat{u}_i$ it is possible that some embedding $\hat{v}_j$ is needed multiple times in the call stack. In this case, $\hat{v}_j$ and all of its dependencies will be generated multiple times. To avoid repeated computation, we cache the result of any predicted embedding computation and return it instead of recomputing. These embedding caches are wiped after every gradient update, since they no longer reflect what the model would have generated. This optimization reduces the number of requests for embeddings by two orders of magnitude. 

\textbf{Evidence Limit (EL)} 
On larger datasets such as ML-10M, a single user or item can have thousands of ratings. This means that generating one embedding may require the intermediate generation of thousands of other embeddings, a costly procedure at each level of our recursive algorithm.  Instead, we define an evidence limit $\el \in \mathbb{N}$ . When generating an embedding we randomly select an $\el$ number of ratings and use them to generate our embedding. We set $\el$ to 80 for all of our experiments unless otherwise specified. Using $\el$ roughly halves the number of embeddings we generate on ML-100K, with the gains increasing on larger datasets.

\textbf{Prototype Prioritization (PP)} 
Once an evidence limit is introduced, picking the right embeddings to explore becomes an optimization problem. In the worst case, every single embedding we select is unable to reach the prototype section by the time Max Depth is reached, and will therefore return None. To avoid this where possible, we use Prototype Prioritization. Instead of randomly sampling $\el$ users, we greedily select available prototypes to generate our embedding.  If the number of available prototypes is less than $\el$, we randomly sample the rest to ensure that we still generate from $\el$ embeddings.  On ML-100K this reduces the number of generated embeddings by 36\%.

\textbf{Telescoping Evidence Limit (TEL)} The deeper into the recursion stack an embedding is generated, the weaker the evidence it provides about the requested rating. To help reduce the amount of distantly-related embeddings we collect as evidence, we use a Telescoping Evidence Limit. The $\textbf{TEL}$ is a depth aware version of $\el$, which halves its value at each depth. Thus $\textbf{TEL}^{d=0} \coloneqq \el$ and in general $\textbf{TEL}^{d=i} \coloneqq \frac{\el}{2^i}$.

\section{Experiments} \label{sec:experiments}

\textbf{Datasets.}
We evaluate our model on three collaborative filtering datasets: MovieLens-100K, MovieLens-1M and MovieLens-10M~\cite{harper2016movielens}. The ML-100K dataset contains 100,000 ratings of 1682 movies by 943 users,  ML-1M dataset contains approximately 1 million ratings of 3900 movies by 6040 users and ML-10M dataset contains approximately 10 million ratings of 10681 movies by 71567 users. For our experiments, we take a 80/20 train/validation split of the datasets. Our hyperparameters are optimized according to this choice of validation set. 

\textbf{Implementation details.}
The standard configuration we took for REC, unless otherwise specified, is as follows: the number of prototypical users and items $\abs{\pu}$ and $\abs{\pv}$ are set to 50, the Evidence Limit $\el$ and Max Depth $\md$ are set to 80 and 4 respectively. We use a 3-layer feed-forward neural network with each 200 neurons for each hidden layer. The activation functions of each net are ReLUs, with exception of the last layer which is taken to be linear. 

In addition, we introduce a pretraining phase in our REC experimentation: we first perform PMF on only the prototypical block before training on all parameters. This procedure ensures that the prototypes are large enough to recombine into an accurate rating, and therefore reducing the need for updating its distribution. We found that this short constant-time procedure sped up the optimization process considerably. Lastly, we set the default batch size to be 1000 and use the Adam optimizer~\cite{Adam:14}  with a learning rate of $10^{-3}$ and regularization parameter $\lambda=10^{-5}$. Our metric of evaluation across all experiments is test root-mean square error (RMSE). 

\textbf{Test RMSE comparisons.}
We begin by comparing the performance of REC to the following collaborative-filtering algorithms: PMF~\cite{mnih2008probabilistic}, NNMF~\cite{dziugaite2015neural}, Biased-MF~\cite{koren2009matrix}, and CF-NADE~\cite{zheng2016neural}. For ML-100K and ML-1M, we use the standard configuration whereas for ML-10M, the only changes we make are setting $\el$ to be 40 and $\md$ to be 3. In all three experiments, we train for 2000 iterations. Table~\ref{RMSE comparison table} gives the comparison scores. We find that for ML-100K, REC achieves a test RMSE performance comparable to standard collaborative-filtering algorithms. For ML-1M and ML-10M, while our method does not reach state-of-art performance, especially compared to CF-NADE and BiasedMF, we believe that this can be substantially improved if we tune the number of prototype users $\pu$ and items $\pv$. 


\begin{table}[h]
  \caption{Test RMSE results on ML-100K, ML-1M, and ML-10M for various models. Scores reported for PMF, NNMF, Biased-MF (ML-100K/ML-1M) were taken from~\cite{dziugaite2015neural}. Scores reported for CF-NADE and Biased-MF (ML-10M) were taken from~\cite{zheng2016neural}. Note that these results were obtained using a 90/10 train/valid split whereas in REC we used a 80/20 split.} 
  \label{RMSE comparison table}
  \centering
  \begin{tabular}{lllll}
    \toprule
   Model     &  ML-100K     & ML-1M & ML-10M \\
    \midrule
    PMF~\cite{mnih2008probabilistic} & 0.952  & 0.883   & -  \\
    NNMF~\cite{dziugaite2015neural}     & 0.903 & 0.843  &  -   \\
    Biased-MF~\cite{koren2009matrix} & 0.911 & 0.852 & 0.803 \\
    CF-NADE~\cite{zheng2016neural} & - & 0.829 & 0.771 \\
    REC     &     0.910   & 0.882 & 0.846 \\
    \bottomrule
  \end{tabular}
\end{table}

\subsection{Coupling of parameters leads to faster convergence} \label{subsec:fast training}

We compare the performance of REC to PMF in the early stages of the training process. Again, we use the standard configuration for REC on ML-100K but change $\el$ and $\md$ to 20 and 2 for ML-1M and ML-10M. For PMF, the same 80/20 train/valid split is used to partition the datasets and we choose batch-sizes of 1000, 5000, and 10000 for ML-100K, ML-1M, and ML-10M respectively. The reason behind increasing batch-sizes is that for PMF, as opposed to REC, larger batch-sizes are required to achieve accurate training as the dataset size grows. For consistency, we also applied the same pretraining procedure in REC to PMF. 

\begin{figure}[h]
\begin{subfigure}[t]{\textwidth}
    \centering
      \includegraphics[scale=.5]{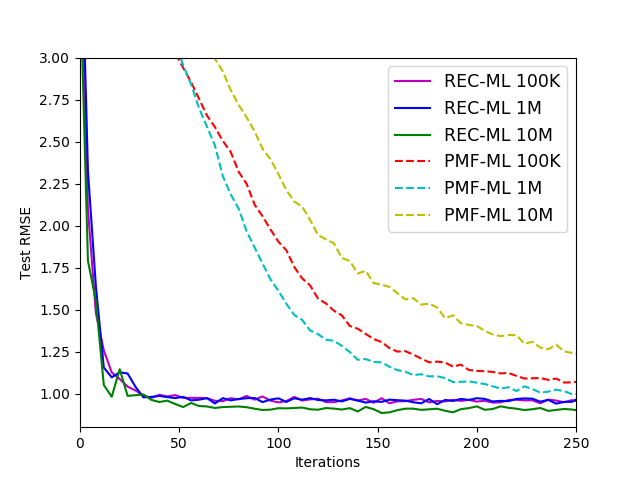}
\end{subfigure}%

\caption{Performance of REC and PMF on ML-100K, ML-1M and ML-10M for $<250$ iterations.}
\label{fig:fastiterations}
\end{figure}

In Figure~\ref{fig:fastiterations}, we see that for all three datasets, REC converges to a RMSE $<1$ in under 30 iterations. Furthermore, we find that REC learns very well in the early training process. On the other hand, PMF cannot reach a RMSE $<1$ in $~250$ iterations; in fact, it only attains this at around iterations 400, 360, and 500 for ML-100K, ML-1M, and ML-10M respectively. As for wall-clock statistics, it takes around 10, 45 and 70 seconds to complete one REC iteration on ML-100K, ML-1M and ML-10M respectively. 

As demonstrated in Figure~\ref{fig:fastiterations}, REC has similar test RMSE convergence curves across increasingly large datasets while maintaining a constant number of parameters.  This highlights the attractive scalability properties of REC.  In contrast, we observe that the number of iterations it takes for PMF to converge depends on the size of the dataset.

\subsection{Constant Scale Online Predictions Without Retraining}



\begin{figure}
    \centering%
    \begin{subfigure}[t]{.45\textwidth}%
    \centering%
     \includegraphics[width=\linewidth]{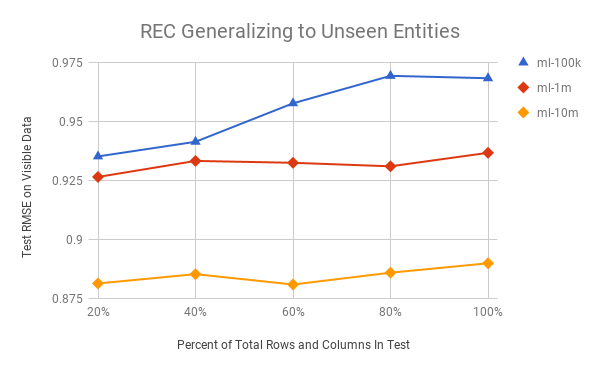}
    \caption{REC's incremental performance on the test set as a function of percentage of total rows and columns available.}
    \label{fig:online-rec}
    \end{subfigure}%
    \hfill
\begin{subfigure}[t]{.45\textwidth}
    \centering%
    \includegraphics[width=\linewidth]{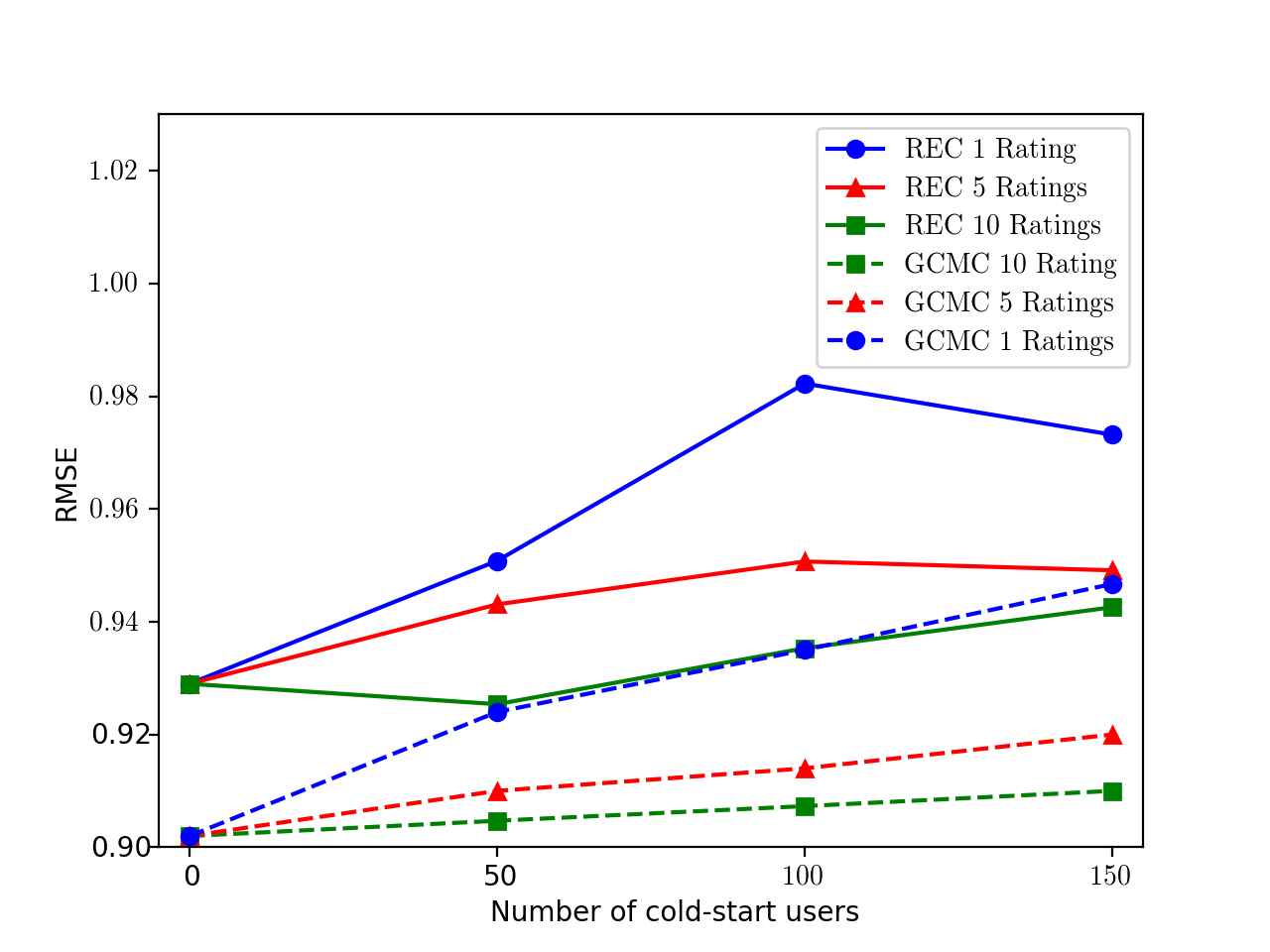}
    \label{fig:cold-start}
    \caption{Cold start analysis for REC and GC-MC where $N_c \in [0, 50, 100, 150]$ and $n_r \in [1, 5, 10]$.  Scores reported for GC-MC were taken from~\cite{BKW:17}}
    \end{subfigure}

\caption{REC's performance to handle two sparsity problems: online learning and cold start}
\label{fig:REC performance plots}
\end{figure}

Here, we evaluate REC's performance on an online learning problem. We do this by training REC on a small subset of all rows and columns, and then testing its performance against increasingly large sets of new rows and columns.
This approximates new content and subscribers being added to a recommender system over time. To keep our experimental setup simple, we assume that we have access to the entire dataset for the purpose of determining our prototypes. We begin by selecting 50 prototypes for both users and items. We then train to convergence on 20\% of the rows and columns of our dataset where this 20\% includes the prototypes. We reintroduce 20\% of the total rows and columns at a time to both the train and test sets, and evaluate REC's performance on the test set at each iteration.  The incremental findings of this experiment can be seen in Figure~\ref{fig:online-rec} while the final results are showcased in Table~\ref{rmse table growth}.

\begin{table}[h]
  \caption{Online test results on ML-100K, ML-1M, and ML-10M. Recall that the number of prototype users $\pu$ and prototype items $\pv$ are fixed to 50 for all datasets.} 
  \label{RMSE comparison table}
  \centering
  \begin{tabular}{lllll}
    \toprule
   Model     &  ML-100K     & ML-1M & ML-10M \\
    \midrule
    Test RMSE on full novel dataset & 0.967  & 0.936  & 0.889  \\
    Parameters for REC (in millions)     & .17 & .17  &  .17   \\
    Percent of data seen in training & 6.02\% & 6.2\% & 6.03\% \\
    Number of new rows and columns in test set & 2020 & 7951 & 65798 \\
    \bottomrule
  \end{tabular}
  \label{rmse table growth}
 \end{table}

\subsection{Cold Start}

In this experiment on ML-100K, we show that \coolname[] generates accurate predictions for users with few ratings. 
To simulate a cold start setting, we utilize the formulation in~\cite{BKW:17}. Out of our training set, we randomly select $N_c$ users to be our cold start users, by dropping all but $n_r$ of their ratings. After training, we log REC's performance on the entire validation set.  In order to study the effect of both $N_c$ and $n_r$ on our model, we report configurations of $N_c \in [0, 50, 100, 150]$ and $n_r \in [1, 5, 10]$.  Note that the case where $N_c = 0$ is regular REC, as no users have ratings dropped. While REC underperforms when compared to the results given in~\cite{BKW:17}, it significantly outperforms guessing the mean (1.15 RMSE), and our experiments using REC exhibit similar trends to those given in~\cite{BKW:17}, indicating that REC is a promising technique for address cold-start problems.

\section{Related Work} \label{sec:related work}
Incorporating deep learning techniques into collaborative filtering has been an active area of research~\cite{ZYS:17}. In~\cite{dziugaite2015neural}, the authors introduced Neural Network Matrix Factorization (NNMF)~\cite{dziugaite2015neural} which uses a neural network to factorize the ratings matrix $\c{R}$. CF-NADE~\cite{zheng2016neural} is a neural autoregressive architecture for collaborative filtering inspired from Restricted Boltzmann Machine for Collaborative Filtering (RBM-CF)~\cite{salakhutdinov2007restricted} and Neural Autoregressive Distribution Estimator (NADE)~\cite{larochelle2011neural}, currently maintains the highest state-of-the-art performance across MovieLens and Netflix datasets, though with high complexity overhead. Neural Collaborative Filtering (NCF)~\cite{he2017neural} provides a systematic study of applying various neural network architectures into the collaborative filtering problem.

Graph Convolutional Matrix Completion (GC-MC)~\cite{BKW:17}, similar to REC, also interpreted matrix completion problem as a bipartite user-item graph where observed ratings represent links. A graph convolutional auto-encoder framework was used in GC-MC to predict the links. The main difference is that GC-MC stores the latents of every user and item, whereas REC only stores a small subset, generating the rest through neural networks.

Online collaborative filtering methods have also been an active vein of research. The authors in~\cite{abernethy2007online} proposed an algorithm for learning a rank-$k$ matrix factor model in an online manner, which scales linearly with $k$ and the number of ratings. An online algorithm to learn rank-prediction rules for a user, using the ratings of other users, was proposed in the work of~\cite{crammer2002pranking}. In the work of~\cite{bresler2014latent}, the authors present an algorithm that learns to group users into one of $k$ types of users in an online fashion, where each of the $k$ user types have their own established probabilities of liking each item.

\section{Conclusion} \label{sec:conclusion}
In this paper, we proposed REC, a generalization of rowless and columnless matrix factorization techniques where user and item embeddings are generated through recursive evidence chains. Our model has a variety of interesting and attractive properties, such as constant parameter scaling, fast training, and the ability to handle both online learning and the cold start problem. We demonstrate its performance on standard datasets and find that it has competitive performance to existing collaborative-filtering algorithms.

\bibliographystyle{plain}
\bibliography{main}

\begin{thebibliography}{10}

\bibitem{abernethy2007online}
Jacob Abernethy, Kevin Canini, John Langford, and Alex Simma.
\newblock Online collaborative filtering.
\newblock {\em University of California at Berkeley, Tech. Rep}, 2007.

\bibitem{bresler2014latent}
Guy Bresler, George~H Chen, and Devavrat Shah.
\newblock A latent source model for online collaborative filtering.
\newblock In {\em Advances in Neural Information Processing Systems}, pages
  3347--3355, 2014.

\bibitem{crammer2002pranking}
Koby Crammer and Yoram Singer.
\newblock Pranking with ranking.
\newblock In {\em Advances in neural information processing systems}, pages
  641--647, 2002.

\bibitem{das2016chains}
Rajarshi Das, Arvind Neelakantan, David Belanger, and Andrew McCallum.
\newblock Chains of reasoning over entities, relations, and text using
  recurrent neural networks.
\newblock {\em arXiv preprint arXiv:1607.01426}, 2016.

\bibitem{dziugaite2015neural}
Gintare~Karolina Dziugaite and Daniel~M Roy.
\newblock Neural network matrix factorization.
\newblock {\em arXiv preprint arXiv:1511.06443}, 2015.

\bibitem{harper2016movielens}
F~Maxwell Harper and Joseph~A Konstan.
\newblock The movielens datasets: History and context.
\newblock {\em ACM Transactions on Interactive Intelligent Systems (TiiS)},
  5(4):19, 2016.

\bibitem{he2017neural}
Xiangnan He, Lizi Liao, Hanwang Zhang, Liqiang Nie, Xia Hu, and Tat-Seng Chua.
\newblock Neural collaborative filtering.
\newblock In {\em Proceedings of the 26th International Conference on World
  Wide Web}, pages 173--182. International World Wide Web Conferences Steering
  Committee, 2017.

\bibitem{Adam:14}
D.~P. {Kingma} and J.~{Ba}.
\newblock {Adam: A Method for Stochastic Optimization}.
\newblock {\em ArXiv e-prints}, December 2014.

\bibitem{koren2009matrix}
Yehuda Koren, Robert Bell, and Chris Volinsky.
\newblock Matrix factorization techniques for recommender systems.
\newblock {\em Computer}, 42(8), 2009.

\bibitem{larochelle2011neural}
Hugo Larochelle and Iain Murray.
\newblock The neural autoregressive distribution estimator.
\newblock In {\em Proceedings of the Fourteenth International Conference on
  Artificial Intelligence and Statistics}, pages 29--37, 2011.

\bibitem{mnih2008probabilistic}
Andriy Mnih and Ruslan~R Salakhutdinov.
\newblock Probabilistic matrix factorization.
\newblock In {\em Advances in neural information processing systems}, pages
  1257--1264, 2008.

\bibitem{rennie2005fast}
Jasson~DM Rennie and Nathan Srebro.
\newblock Fast maximum margin matrix factorization for collaborative
  prediction.
\newblock In {\em Proceedings of the 22nd international conference on Machine
  learning}, pages 713--719. ACM, 2005.

\bibitem{salakhutdinov2007restricted}
Ruslan Salakhutdinov, Andriy Mnih, and Geoffrey Hinton.
\newblock Restricted boltzmann machines for collaborative filtering.
\newblock In {\em Proceedings of the 24th international conference on Machine
  learning}, pages 791--798. ACM, 2007.

\bibitem{srebro2003weighted}
Nathan Srebro and Tommi Jaakkola.
\newblock Weighted low-rank approximations.
\newblock In {\em Proceedings of the 20th International Conference on Machine
  Learning (ICML-03)}, pages 720--727, 2003.

\bibitem{BKW:17}
R.~{van den Berg}, T.~N. {Kipf}, and M.~{Welling}.
\newblock {Graph Convolutional Matrix Completion}.
\newblock {\em ArXiv e-prints}, June 2017.

\bibitem{verga2016generalizing}
Patrick Verga, Arvind Neelakantan, and Andrew McCallum.
\newblock Generalizing to unseen entities and entity pairs with row-less
  universal schema.
\newblock {\em arXiv preprint arXiv:1606.05804}, 2016.

\bibitem{ZYS:17}
S.~{Zhang}, L.~{Yao}, and A.~{Sun}.
\newblock {Deep Learning based Recommender System: A Survey and New
  Perspectives}.
\newblock {\em ArXiv e-prints}, July 2017.

\bibitem{zheng2016neural}
Yin Zheng, Bangsheng Tang, Wenkui Ding, and Hanning Zhou.
\newblock A neural autoregressive approach to collaborative filtering.
\newblock In {\em Proceedings of the 33nd International Conference on Machine
  Learning}, pages 764--773, 2016.

\end{thebibliography}

\end{document}